\documentclass[prb,twocolumn,reprint,superscriptaddress,showpacs]{revtex4-1}

\usepackage[version=3]{mhchem} 
\usepackage{amssymb} 
\usepackage{textcomp} 

\usepackage[utf8]{inputenc}

\begin{document}

\author{C. M. Polley}
\email{craig.polley@maxlab.lu.se}
\affiliation{MAX IV Laboratory, Lund University, 221 00 Lund, Sweden}
\author{P. Dziawa}
\author{A. Reszka}
\author{A. Szczerbakow}
\author{R. Minikayev}
\author{J. Z. Domagala}
\author{S. Safaei}
\author{P. Kacman}
\author{R. Buczko}
\affiliation{Institute of Physics, Polish Academy of Sciences, 02-668 Warsaw, Poland}
\author{J. Adell}
\affiliation{MAX IV Laboratory, Lund University, 221 00 Lund, Sweden}
\author{M. H. Berntsen}
\altaffiliation{Present address: Deutsches Elektronen-Synchrotron (DESY), Photon Science, Coherent X-ray Scattering, Notkestrasse 85, 22607 Hamburg, Germany}
\author{B. M. Wojek}
\author{O. Tjernberg}
\affiliation{KTH Royal Institute of Technology, ICT Materials Physics, Electrum 229, 164 40 Kista, Sweden}
\author{B. J. Kowalski}
\author{T. Story}
\affiliation{Institute of Physics, Polish Academy of Sciences, 02-668 Warsaw, Poland}
\author{T. Balasubramanian}
\affiliation{MAX IV Laboratory, Lund University, 221 00 Lund, Sweden}

\title{Observation of topological crystalline insulator surface states on (111)-oriented Pb$_{1-x}$Sn$_x$Se films}

\begin{abstract}
We present angle resolved photoemission spectroscopy measurements of the surface states on \textit{in-situ} grown (111) oriented films of Pb$_{1-x}$Sn$_{x}$Se, a three dimensional topological crystalline insulator. We observe surface states with Dirac-like dispersion at $\bar{\Gamma}$ and $\bar{M}$ in the surface Brillouin zone, supporting recent theoretical predictions for this family of materials. We study the parallel dispersion isotropy and Dirac-point binding energy of the surface states, and perform tight-binding calculations to support our findings. The relative simplicity of the growth technique is encouraging, and suggests a clear path for future investigations into the role of strain, vicinality and alternative surface orientations in (Pb,Sn)Se compounds.
\end{abstract}


\maketitle
\section{Introduction}
Following the first experimental report in 2008 of photoemission from the unusual surface states associated with a three dimensional topological insulator (TI) \cite{Hsieh2008}, these materials have attracted a rapidly developing research interest \cite{Ando2013}. More than simply fertile new ground for research into quantum phenomena, topological insulators have much to offer in spintronic and quantum computation applications due to their spin-polarized, topologically protected interface states.

For the initially studied class of Z$_2$ invariant topological insulators, the gapless nature of these surface states is topologically protected by time reversal symmetry. However it was recently realized that crystal symmetry can play a similar role, resulting in the new class of `topological crystalline insulators' (TCI)\cite{Fu2011, Hsieh2012}. To date TCI surface states have been experimentally observed on the (100) faces of Pb$_{1-x}$Sn$_x$Se \cite{Dziawa2012, Wojek2013, Gyneis2013, Okada2013}, Pb$_{1-x}$Sn$_x$Te \cite{Xu2012, Tanaka2013} and SnTe \cite{Tanaka2012, Sadfar2013}. 

For all three materials, angle resolved photoemission (ARPES) studies of the (100) facet show two spin-polarized Dirac-like surface states close to each $\bar{X}$ point in the surface Brillouin zone (SBZ). The nontrivial surface states derive from the inverted bulk band gap at the $L$-points of the bulk Brillouin zone, and are protected by \{011\} mirror planes. The projection of two inequivalent $L$-points to the same $\bar{X}$ point in the SBZ gives rise to a complex Fermi surface which exhibits a Lifshitz transition as a function of the chemical potential \cite{Hsieh2012}.

The fundamental role of crystalline symmetry in this new family of topological materials makes the study of different surface orientations attractive. Differing degrees of mirror symmetry are retained for different orientations, with important consequences for the low-energy electronic structure. This concept has been discussed at length in recent theoretical studies \cite{Safaei2013,Liu2013} encompassing (100), (110) and (111) surfaces of the (Pb,Sn)Te system. On the (111) facet each bulk $L$-point is projected to a unique, time-reversal invariant momentum in the SBZ ($\bar{\Gamma}$ and each $\bar{M}$). The proposed existence of a symmetrical $\bar{\Gamma}$ surface state with high Fermi velocity and simple spin texture is appealing from the perspective of potential device applications. There is hence strong motivation for an experimental photoemission spectroscopy study of (111) oriented materials, to advance both fundamental and applied aspects of TCI research.

However a practical difficulty encountered when studying non-(100) surface orientations within the (Pb,Sn)Te or (Pb,Sn)Se material classes is the lack of natural cleavage planes. In all previous experimental studies, pristine (100) surfaces could be obtained by \textit{in-situ} cleaving of an \textit{ex-situ} prepared bulk crystal. This method is constrained to the set of natural cleavage planes in the bulk crystal, and hence precludes studying arbitrary surface orientations. Here we report the \textit{in-situ} growth and angle resolved photoemission spectroscopy characterization of (111) oriented Pb$_{1-x}$Sn$_x$Se films. The demonstrated ease of growing such films should widen the scope of future studies in this family of TCI materials.

\section{Epitaxial layer growth}
Samples were grown on BaF$_2$ substrates, which were cleaved in air to expose a fresh (111) facet and then attached to silicon baseplates to enable direct current heating. After entry into ultra high vacuum (UHV), substrates were outgassed at 600~\textcelsius{} for one hour, then reduced to 330~\textcelsius{} for the film growth. Films of Pb$_{1-x}$Sn$_x$Se were deposited onto the BaF$_2$ using an open hot wall epitaxy method\cite{HWE_Review}. Single-source evaporators were used, consisting of filament heated quartz crucibles loaded with crushed, \textit{ex-situ} grown Pb$_{1-x}$Sn$_{x}$Se. Both $x=0.24$ and $x=0.27$ sources were used in the present study. After establishing a crucible temperature of $\approx$330\textcelsius{} (maintaining vacuum pressure below 1$\times$10$^{-8}$~mbar), a substrate was positioned over the crucible mouth for 100 minutes. Subsequent analysis with cross sectional electron microscopy indicates that this corresponds to a film thickness of (900$\pm$50)~nm.
\begin{figure}	
	\includegraphics{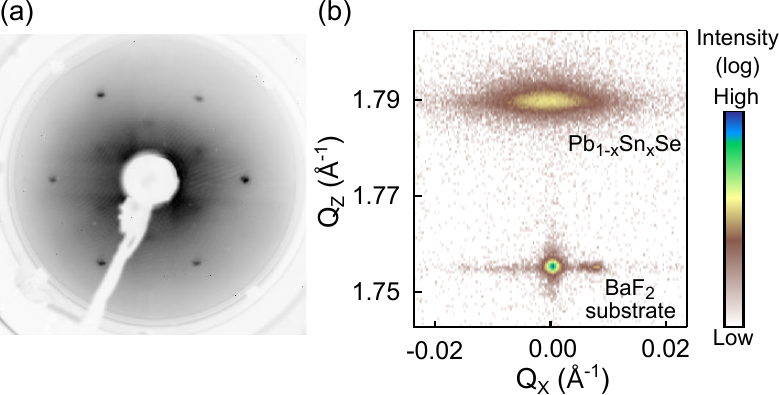}
	\caption{(Color online) (a) Low energy (90~eV) electron diffraction pattern of the grown film, demonstrating first and second order (111) patterns. (b) X-ray reciprocal lattice map using Cu K$\alpha _1$ X-rays, symmetrical to the (111) Bragg peak. The upper feature corresponds to the grown layer and the lower to the BaF$_2$ substrate.
	}
\end{figure}

Structural characterization of the grown layers is summarized in Figure 1. Crucially for the present study, low energy electron diffraction (Fig. 1a) consistently exhibits a (111) pattern, indicating that the grown films have assumed the orientation of the BaF$_2$ substrate. A representative sample grown from the $x=0.27$ source was selected for more comprehensive characterization, including reciprocal space mapping using Cu K$\alpha _1$ X-rays and a high resolution diffractometer, as shown in Figure 1b. Symmetrical reciprocal space maps of the 111 reflection for both the substrate (lower feature) and layer (upper feature) are shown. The grown layer exhibits a higher mosaicity than the substrate, with an angular spread reaching a full width at half maximum of approximately 0.5\textdegree. The shape of the upper pattern demonstrates a well defined lattice constant (6.08~\AA) which can be mapped to a molar fraction of $x=(0.37\pm0.01)$\cite{Szczerbakow1994}. Additional reciprocal space maps (not shown here) of the 115 asymmetric reflection indicate that the grown layer is not strained at room temperature. 

Quantitative composition analysis by energy dispersive X-ray spectroscopy (EDX) indicates a Sn content of $x=(0.36\pm0.01)$, laterally uniform throughout the film and in good agreement with the value calculated from reciprocal space mapping. This is higher than the source material ($x=0.27$) but is still suitable for the observation of a band-inverted condition \cite{Dziawa2012}. We note that in this temperature range evaporation occurs molecularly as PbSe and SnSe \cite{Springholz2007}. Consequently, while the higher vapor pressure of SnSe compared to PbSe may result in small differences between the chemical composition of the source and the grown layer, reasonable preservation of the source composition is possible with a simple, single source evaporation method.

Taken together, the characterization studies demonstrate that epitaxial, compositionally uniform (111) oriented Pb$_{1-x}$Sn$_x$Se films were produced by this growth method, with a Sn content appropriate for the existence of topological surface states. This is essential for the validation of ARPES measurements, which we now discuss. 
 
\section{Photoemission measurements}
 
The film growth was performed on the I4 beam line at the MAX-III synchrotron facility \cite{Jensen1997}, allowing for extensive ARPES characterization without leaving UHV. All spectra were acquired with linearly p-polarized photons at a sample temperature of $\approx$100~K. The photoelectron analyzer was configured for an energy resolution of 25~meV and angular resolution of $\approx0.1^\circ$. Fermi level positions were referenced to a tantalum foil in electrical contact with the samples.

\begin{figure}	
	\includegraphics{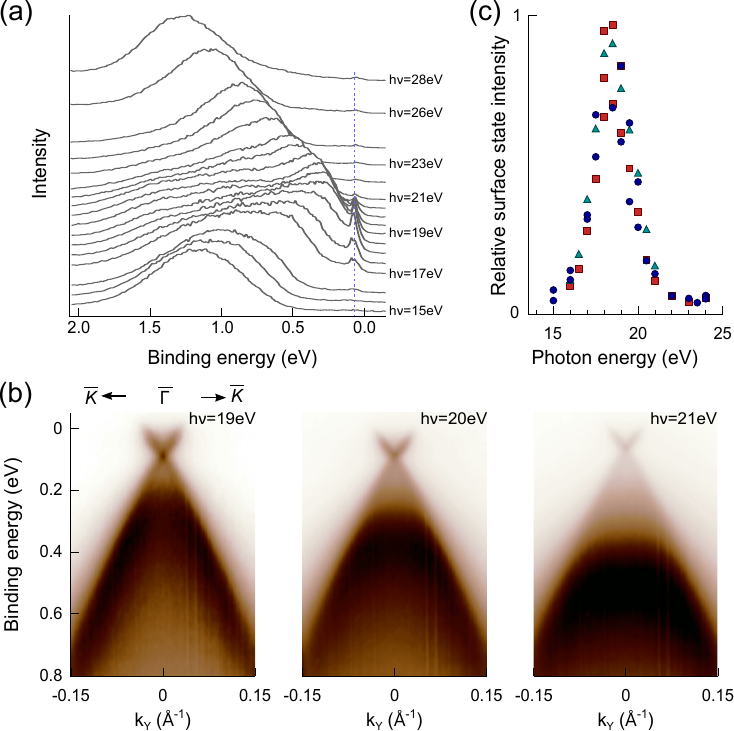}
	\caption{(Color online)  Overview of the role of photon energy in photoemission spectra acquired at $\bar{\Gamma}$. Normal emission energy dispersion curves are plotted in (a), normalized to the same maximum intensity and offset according to the excitation photon energy. A sharp state close to the Fermi level is seen which exhibits no perpendicular momentum dispersion. This is most clearly seen in parallel momentum resolved spectra (b), and signifies that this is a surface state. The intensity of the surface state relative to the nearby bulk band is strongly dependent on the choice of photon energy, shown in (c) for multiple sample preparations (indicated by different markers). }
\end{figure}

The key result of this study is the observation of surface states with Dirac-like dispersion, occurring at the $\bar{\Gamma}$ and the $\bar{M}$ positions in the SBZ. Before studying these states in detail, we first provide confirmation that they are indeed surface states. In Figure 2 we show the results of photon energy dependent ARPES measurements. For the $\bar{\Gamma}$ state seen at normal emission, such a measurement probes the $\Gamma$ - $L$ high symmetry direction in the bulk Brillouin zone. Normal emission energy dispersion curves (Figure 2a) show a broad dispersive peak (attributed to the bulk $L_6$ band) together with a sharper, dispersionless peak at a binding energy of 70~meV. The lack of perpendicular dispersion is more clearly apparent in the parallel momentum resolved spectra shown in Figure 2b, and combined with its position within a bulk bandgap serves to confirm that this second peak originates from a surface state.

While the position of the surface state is unchanged when varying the photon energy, the intensity of the state is strongly modulated. The highest intensity is observed at approximately $h\nu=18.5$~eV; the bulk-state dispersion shown in Figure 2a suggests that this energy probes the $L_6$ valence band maximum. In Figure 2c we illustrate this modulation more clearly by plotting the intensity ratio of the surface state to the bulk $L_6$ band. Similar intensity modulations were observed at approximately 68~eV and 148~eV. Periodic modulation of photoemission intensity with photon energy is a common observation for surface states \cite{Bartynski1985,Hofmann2002,Miwa2013}, with an accepted explanation in terms of a resonant enhancement of the coupling between initial and final states in the photoexcitation process \cite{Louie1980}. The enhancement is strongest for perpendicular momenta which minimize the energy separation between the surface state and bulk band it derives from. In the present case, this implies resonance peaks at photon energies which probe the bulk valence band maxima at $L$ points. A similar resonant enhancement was observed for the $\bar{M}$ surface state, which while significantly weaker than the $\bar{\Gamma}$ state was most intense at a photon energy of 24~eV. 

  \begin{figure}	\includegraphics{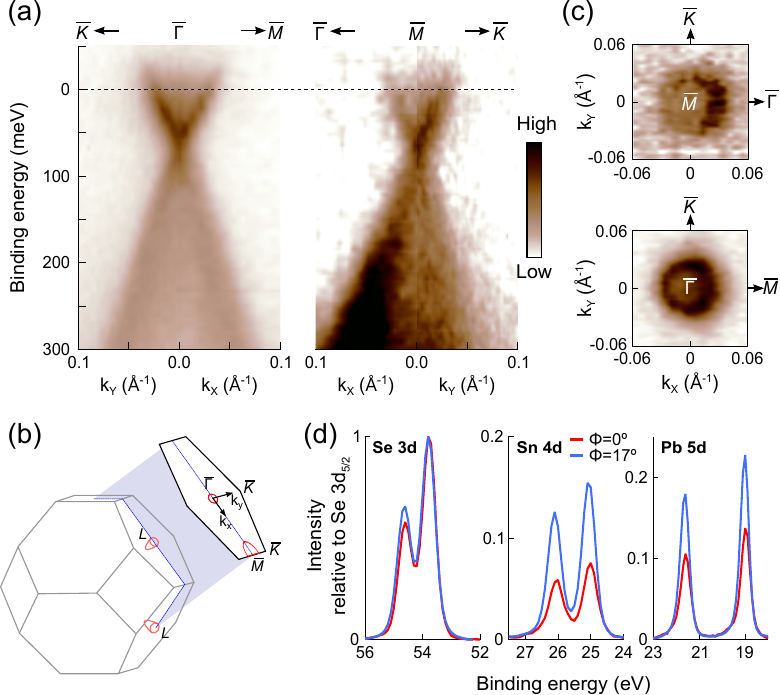}\caption{(Color online) Photoemission spectra of the surface states on a (111) oriented Pb$_{1-x}$Sn$_{x}$Se film. (a) Energy-momentum spectra along high-symmetry lines at $\bar{\Gamma}$ ($k_x=0~\AA^{-1}$) and $\bar{M}$ ($k_x=0.84~\AA^{-1}$) show the presence of Dirac-like surface states at both locations. The location of these states within the SBZ follow from the projection of the bulk $L$-points. These projections exist on \{110\} mirror planes, schematically shown for bulk Fermi ellipsoids in (b). Fermi-surface maps (c) show that the $\bar{\Gamma}$ state is circularly symmetrical, and that the anisotropy of the $\bar{M}$ state is very slight at the Fermi level. Comparing normal ($\phi=0$) with off-normal ($\phi=17^\circ$) emission core-level spectra (d) indicates a cation-rich surface, consistent with the Dirac point residing close to the bulk valence band. Spectra were acquired using photon energies of 17.5~eV ($\bar{\Gamma}$), 24~eV ($\bar{M}$) and 130~eV (core levels)}
  \end{figure}

Figure 3 shows energy-momentum ARPES spectra acquired through $\bar{\Gamma}$ ($k_x=0~\AA^{-1}$) and $\bar{M}$ ($k_x=0.84\AA^{-1}$). Both spectra were measured on a sample grown from Pb$_{0.76}$Sn$_{0.24}$Se source material, from which the highest quality data was obtained. Although the composition was not characterized for this sample, there was no observable difference in the band structure compared to samples grown from the Pb$_{0.73}$Sn$_{0.27}$Se source, suggesting a similar composition. Figure 3a shows slight anisotropy in the $\bar{M}$ surface state dispersion, with faster dispersion along the $\bar{M}-\bar{K}$ cut compared to the $\bar{M}-\bar{\Gamma}$ cut. In contrast, the state at $\bar{\Gamma}$ is isotropic. This can be understood by noting that these surface states derive from the anisotropic bulk bands at the $L$-points, which are projected onto the (111) surface Brillouin zone (Figure 3b). The $\bar{M}$-$\bar{\Gamma}$ direction cuts through the long axis of the bulk constant energy ellipsoids at $L$, giving rise to a slower dispersion. The anisotropy is far less severe than that anticipated for SnTe and Pb$_{0.4}$Sn$_{0.6}$Te \cite{Safaei2013,Liu2013}, and at the Fermi level is barely noticeable (Figure 3c). This is consistent with the reduced eccentricity of the Fermi ellipsoids in PbSe compared to PbTe \cite{Svane2010} and is reproduced by the tight binding calculations we will discuss shortly. The Dirac points in Figure 3a are positioned at a binding energy of $\approx$70~meV, a value found to be similar across all sample preparations.  As anticipated for a (111) surface with cationic termination, the Dirac points sit close to the bulk valence band maxima. Angle dependent core level spectroscopy (Figure 3d) provide supporting evidence that the surface is indeed rich in Sn and Pb cations. Within the experimental resolution, no relative binding energy difference can be observed between the $\bar{\Gamma}$ and $\bar{M}$ Dirac points. This is again quite different to calculations for telluride materials, for which the $\bar{M}$ Dirac point is predicted to be 30~meV (Pb$_{0.4}$Sn$_{0.6}$Te)\cite{Safaei2013} to 45~meV (SnTe)\cite{Liu2013} higher in binding energy than the $\bar{\Gamma}$ Dirac point. We highlight that differences in Dirac point binding energies are highly relevant for the interpretation of low-energy transport measurements \cite{Taskin2013}.

\section{Band structure calculations}

\begin{figure}	
	\includegraphics{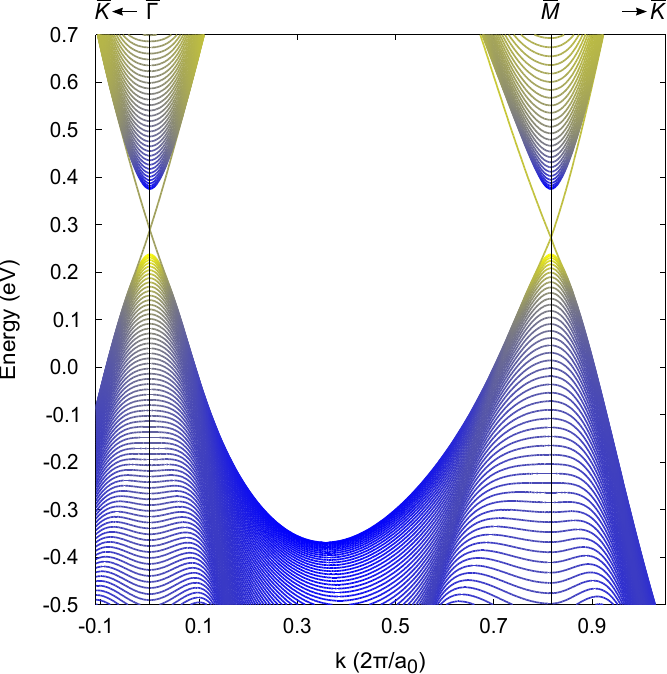}
	\caption{(Color online) The calculated band structure of a (111) oriented and cation terminated Pb$_{0.64}$Sn$_{0.36}$Se slab along high-symmetry directions at $T =$ 100~K ($a_0=6.06$~\AA). The line color denotes the relative contribution of cation (yellow) or anion (blue) p-type orbitals to the wave function.
	}
\end{figure}

To further investigate our experimental characterization of the (111) surface states (in particular the surface state dispersion anisotropy and Dirac point energy positions), we have performed tight-binding (TB) calculations for a (111) oriented cation terminated slab consisting of 451 layers ($\approx$80~nm thick). We have used the virtual crystal approximation for the solid solution of PbSe and SnSe, both in rock-salt structure, using temperature-dependent tight binding parameters described previously \cite{Wojek2013}. We note that the band structure of rock-salt SnSe has not been verified experimentally, and hence the parameterization of this material is based solely on the results of density functional theory calculations. Despite this, the qualitative aspects of the band structure are well captured \cite{Wojek2013}.

As shown in Figure 4, the band calculations for Pb$_{0.64}$Sn$_{0.36}$Se at $T =$ 100~K indicate an inverted bandstructure with Dirac-like surface states at $\bar{\Gamma}$ and $\bar{M}$. Both the dispersion anisotropy and the difference between the $\bar{\Gamma}$ and $\bar{M}$ Dirac-point binding energies are in agreement with the experimental ARPES spectra in Figure 3. The qualitative differences we observe in the surface state properties relative to predictions for (Pb,Sn)Te materials can therefore be anticipated purely from the differences between telluride and selenide bulk properties.

\section{Conclusions}
By growing (111) oriented Pb$_{1-x}$Sn$_x$Se films \textit{in-situ} at a synchrotron ARPES facility, we have been able to spectroscopically measure the topological crystalline insulator states unique to this surface orientation. In contrast to the (100) facet, the Dirac-like surface states are well separated and non-interacting, located at the time reversal invariant momenta $\bar{\Gamma}$ and $\bar{M}$ in the surface Brillouin zone. Our observations are captured by a tight binding model, and provide experimental support for the existing body of theoretical work studying the role of surface orientations in TCI materials. Finally, our demonstration of successfully growing and measuring (Pb,Sn)Se films constitutes an important step towards future studies, enabling investigations into the role of strain \cite{Barone2013} (through lattice-mismatch from different substrates), additional surface orientations (for example the (110) facet) and atomic step density through substrate vicinality.

\textit{Note:} Immediately prior to the submission of this work we became aware of a preprint performing a similar study on (111) oriented SnTe \cite{Tanaka2013_2}.

\begin{acknowledgments}
This work was made possible through support from the Knut and Alice Wallenberg Foundation, the Swedish Research Council, the European Commission Network SemiSpinNet (PITN-GA-2008-215368), the European Regional Development Fund through the Innovative Economy Grant (No. POIG.01.01.02-00-108/09), and the Polish National Science Centre (NCN) Grant No. 2011/03/B/ST3/02659. P.D. and B. J. K. acknowledge support from the Baltic Science Link project coordinated by the Swedish Research Council, VR. We thank Jacek Osiecki for creating data analysis software used in this study.
\end{acknowledgments}

\bibliographystyle{unsrt}

\end{document}